\documentclass[submission,copyright]{eptcs} %

\usepackage{graphicx}
\usepackage{listings}
\usepackage{colortbl}
\usepackage{hhline}
\usepackage[usenames,dvipsnames]{xcolor}
\lstdefinestyle{SMV} {language=C,captionpos=b,tabsize=3,frame=lines,keywordstyle=\color[rgb]{0.6,0,0},morekeywords={MODULE, VAR, ASSIGN, SPEC, init, next, case,esac,IVAR,self,INVARSPEC,CTLSPEC},numbers=left,showtabs=false,morecomment=[l]--,commentstyle=\color{gray},
breaklines=true,stringstyle=\color[rgb]{0.627,0.126,0.941},
basicstyle=\footnotesize\ttfamily}

\lstdefinestyle{SSI} {language=C,captionpos=b,tabsize=3,frame=lines,keywordstyle=\color[rgb]{0.6,0,0},morekeywords={or, then, s, End, c, cr, cn, cfr, cfn, xs, l, f, clear, bpull },numbers=left,showtabs=false,morecomment=[l]--,
breaklines=true,stringstyle=\color[rgb]{0.627,0.126,0.941},
basicstyle=\footnotesize\ttfamily}

\lstdefinestyle{SCA} {language=C,captionpos=b,tabsize=3,frame=lines,keywordstyle=\color[rgb]{0.6,0,0},morekeywords={var, val, foreach, for, match},numbers=left,showtabs=false,morecomment=[l]//,commentstyle=\color{gray},
breaklines=true,stringstyle=\color[rgb]{0.627,0.126,0.941},
basicstyle=\footnotesize\ttfamily}

\graphicspath{{./pictures/}}
\usepackage{enumerate}

\usepackage{pgfplots}

\usepackage{url}

\newenvironment{keywords}{
       \list{}{\advance\topsep by0.35cm\relax\small
       \leftmargin=1cm
       \labelwidth=0.35cm
       \listparindent=0.35cm
       \itemindent\listparindent
       \rightmargin\leftmargin}\item[\hskip\labelsep
                                     \bfseries Keywords:]}
     {\endlist}
     
\title{Verification of railway interlocking systems}
\author{Simon Busard, Quentin Cappart, Christophe Limbr\'{e}e,\\
		  Charles Pecheur, Pierre Schaus
	\thanks{This research is financed by the Walloon Region as part of the Logistics in Wallonia competitiveness pole.}
	\institute{Universit\'{e} catholique de Louvain, Louvain-La-Neuve, Belgium}
\email{\{simon.busard|quentin.cappart|charles.pecheur|pierre.schaus\}@uclouvain.be}
\email{christophe.limbree@student.uclouvain.be}}

\def\titlerunning{Verification of railway interlocking systems}
\def\authorrunning{S. Busard, Q. Cappart, C. Limbr\'{e}e, C. Pecheur, P. Schaus}

\begin{document}

\maketitle

\begin{abstract}
In the railway domain, an interlocking is a computerised system that controls the railway signalling objects in order to allow a safe operation of the train traffic. Each interlocking makes use of particular data, called application data, that reflects the track layout of the station under control.  The verification and validation of the application data are performed manually and is thus error-prone and costly. In this paper, we explain how we built an executable model in NuSMV of a railway interlocking based on the application data. We also detail the tool that we have developed in order to translate the application data into our model automatically. Finally we show how we could verify a realistic set of safety properties on a real-size station model by customizing the existing model-checking algorithm with PyNuSMV a Python library based on NuSMV.
\end{abstract}


\section{Introduction}
\label{context}

In the railway domain, an $interlocking$ is an arrangement of systems that prevents conflicting train movements in the stations. It is more specially a signalling subsystem that controls the routes, the points and the signals before allowing a train through a station.  Computer-based interlockings are configured based on a set of \emph{application data} particular to each station.  In this paper, the format considered for the application data is the SSI language~\cite{N20} that is the electronic interlocking used by the Belgian railways since 1992.

The safety of the train traffic relies on the correctness of the application data. The \emph{European Railway Agency}\footnote{\url{www.era.europa.eu}} has edited norms in an effort to harmonize the signalling principles and rules at the European level~\cite{CENELEC2011,EN50128}. Those norms strongly recommend the use of formal methods.

\label{manTests}

Currently, the application data are prepared manually and are thus subject to human errors. For example, some prerequisite to the clearance (e.g. green light) of the home signal of a route can be missing. This kind of error can easily be discovered by a code review or by testing on a simulator. However, errors caused by concurrent actions (e.g. route commands) are much harder to find. In this case, the combination of possible concurrent actions explodes quickly and testing all possible combinations manually is impracticable. 

As testing all the possible scenarios is impossible, the manual validation of the application data relies on a relaxed verification process:

\begin{enumerate}
\item The functional tests ensure that the system responds properly to the commands issued by the controller. Those tests are performed by the expert who wrote the application data. 
\item The safety tests check that each command (e.g. route) is tested and all the conditions that are supposed to impact the command are tested in all their possible values. Those tests are prepared and carried out by an independent tester.
\item The application data are reviewed by the engineer in charge of the project.
\end{enumerate}

During this process, all anomalies are traced in a bug management tool and must be fixed before the interlocking is commissioned.

\subsection{Verification of interlockings using model checking}

Our approach to improve the manual verification method is to automatically convert the application data into a model and to verify safety properties on that model with a model checker. A model checker is a tool that automatically checks whether a system meets a given property by comparing the reachable states space of the model of the system and the property.  In our case, we used the NuSMV~\cite{NuSMV2} symbolic model checker for which our research team\footnote{\url{http://lvl.info.ucl.ac.be}} has a broad experience.  We also used PyNuSMV, a Python library based on NuSMV that can be used to prototype new model-checking algorithms~\cite{Busard-Pecheur-13}. PyNuSMV gathers the flexibility of Python and the functionalities of NuSMV in order to efficiently manipulate the BDD data structures.

The safety properties are written based on the track layout of each station and on the safety rules applicable in the signalling domain.  
 
\par Our approach is divided into the following steps:

\begin{enumerate}
\item Generate a model of the interlocking based on the application data.  This is done by a translator tool.
\item Generate a model of the trains using a Domain Specific Language that encodes the track layout.
\item State all the properties that must be verified to ensure safety based on the track layout.
\item Combine the models of the interlocking, of the trains, and of the properties into an SMV model that can be processed by NuSMV.
\item Use specific model-checking procedures developed with PyNuSMV to reduce the execution time and produce additional data (route compatibility tables).
\end{enumerate}

\begin{figure}[ht]
\centering
\includegraphics[width=1\linewidth]{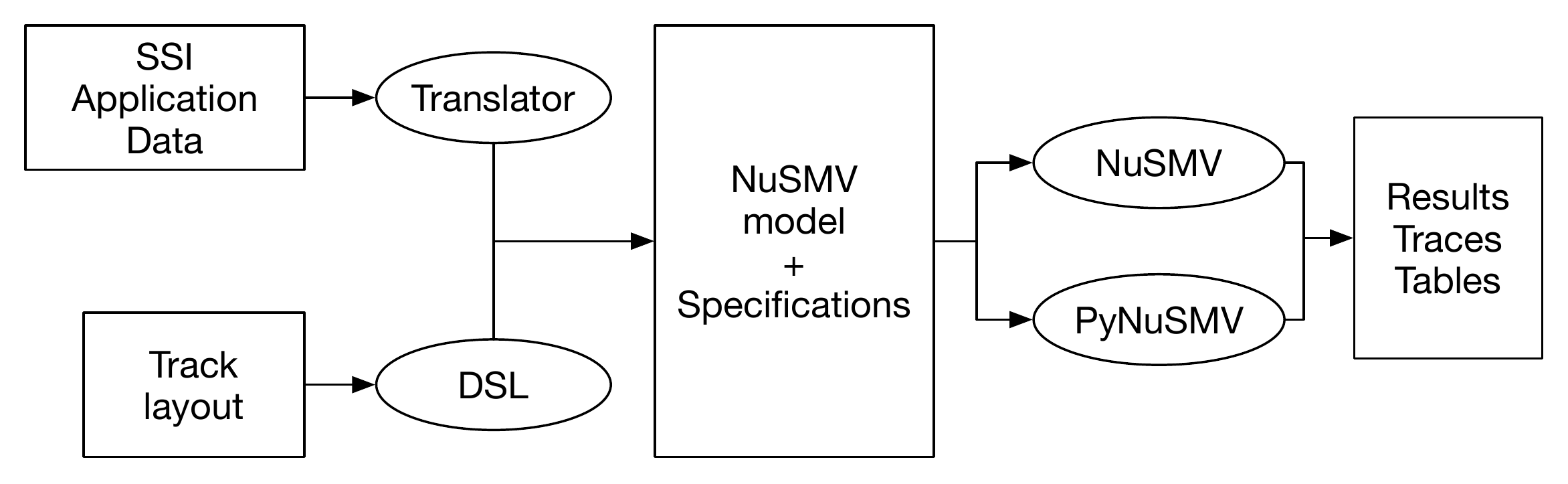} 
\caption{Steps of our approach.}
\label{fig:process} 
\end{figure}

This process is shown in Figure~\ref{fig:process}. Our approach is currently only applicable to a single interlocking. Our other assumptions and abstractions are explained in Section~\ref{assums}.

In the next section, we describe the different components used in our model in order to present our model in Section~\ref{model}. In Section~\ref{translation}, we explain how our model is constructed based on the application data. In Section~\ref{safProp}, we detail the safety properties verified by our model. In Section~\ref{verification}, we discuss how we can improve the performance of the verification. References to related work are provided in Section~\ref{relWork}. 

\section{Interlocking components}
\label{compDesc}

\begin{figure}[ht]
\centering
\includegraphics[width=1\linewidth]{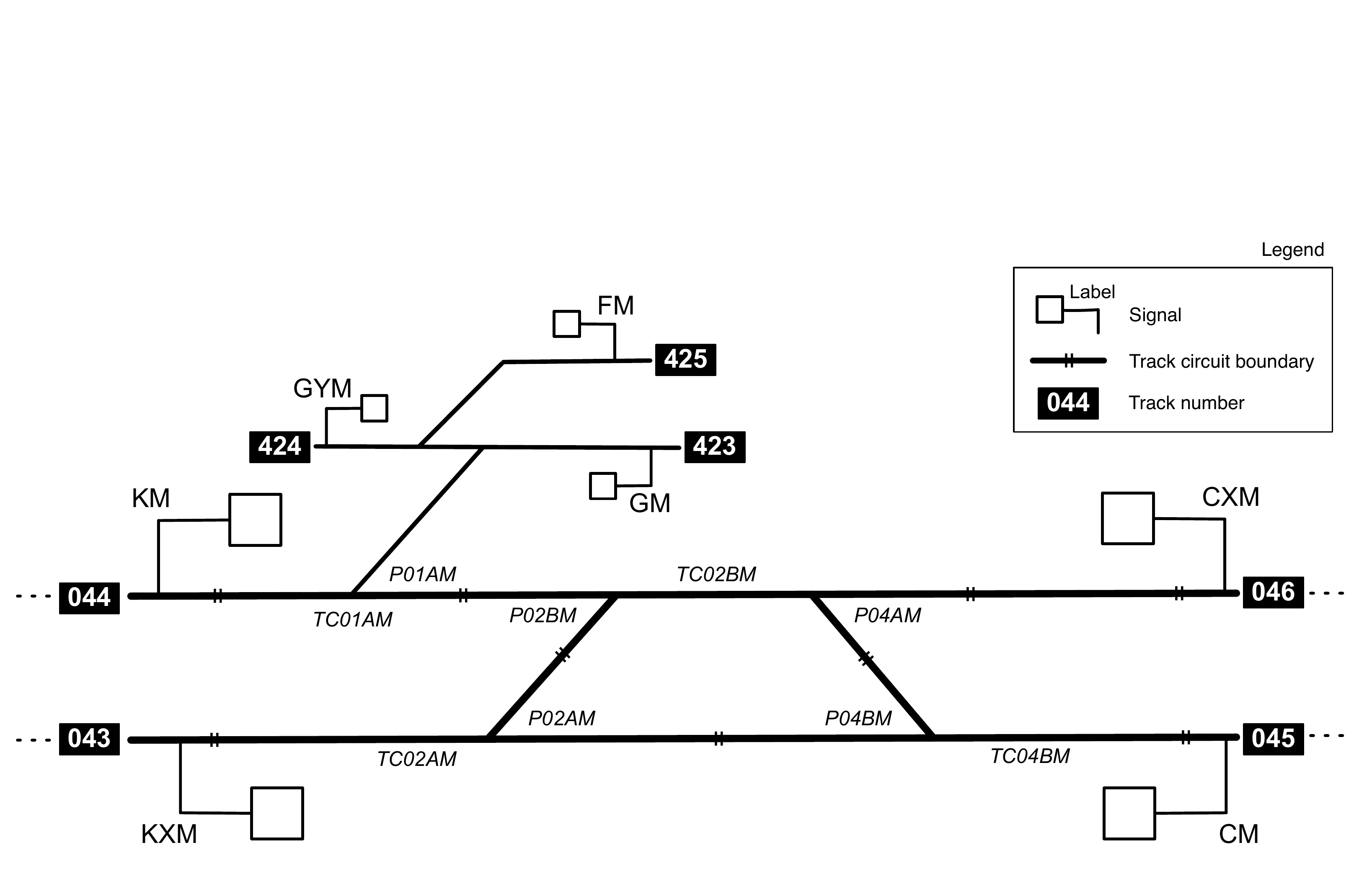} 
\caption{Layout of the Nam\^eche station.}
\label{fig:layout} 
\end{figure}

Figure~\ref{fig:layout} shows the track layout of the station of \emph{Nam\^eche}, a Belgian town. 
This station will be our case study for explaining our approach.
The whole station is controlled by a single interlocking that controls 14 routes.

On this figure, the following elements can be identified:
\begin{itemize}
\item The track identifiers (e.g. 045). 
\item The signals (e.g. KM) that are used to grant access to the routes for the trains.
\item The points (e.g. P02AM) that are the railway junctions allowing a train to move from one track to another.
\item The track circuits\footnote{Track circuits are sometimes called track segments.} (e.g. TC01AM) that are used to detect the vacancy of a portion of the track layout. 

\end{itemize}

The interlocking allows a safe train operation on a railway network or in a station by controlling the \emph{routes}. The routes are the paths followed by the trains when running through a station. For instance, R\_KM\_045 is a route going from signal KM to track 045. The interlocking handles a route command in the following manner:

\begin{enumerate}
\item When a route is requested, it verifies whether the command is safe.  This means that the track components (points and track circuits) requested should not be already reserved for another route (the points P01AM, P02BM, P04AM, P04BM, and the tracks TC01AM, TC02BM, TC04BM for R\_KM\_045).
\item It commands the points by controlling their actuators (points P01AM, P02BM, P04AM, P04BM to the right positions for R\_KM\_045).
\item It verifies the new status of the points by comparing the command and the replied status of the actuators.
\item It then grants access to the train on the route, setting the origin signal of the route to green (KM for R\_KM\_045).
\end{enumerate}

%
%

A route is composed of several segments called subroutes, corresponding to its track segments (three for route R\_KM\_045). Each of them is locked when the route is set and is released when the train has fully freed the home track circuit of the subroute, releasing the corresponding points. 

This process also makes use of other logical components not shown in Figure~\ref{fig:layout} like the component materialising a point locking ($UIR$) or the component recording the train passage on the route ($TISP$).
The list of controls and verifications stated above are loaded from the application data and used by the interlocking for every route. The fact that the application data properly reflects the track layout and the signalling principles is thus crucial in the safety that the interlocking can achieve. That is why so much effort is devoted to their verification.

\section{Model description}
\label{model}

In this section, we describe how the model is designed in order to verify the application data. The complete model can be downloaded from url: \url{http://lvl.info.ucl.ac.be/Tools/InterlockingModel}.

In order to reduce the size of the state space, several assumptions and abstractions were made: 

\label{assums}
\begin{enumerate}

\item Our method is only applicable to areas controlled by a single interlocking. The case of interlockings interconnected in a network will be studied in our future works.
\item Only two signal aspects are modelled: proceed (green), and danger (red). The trains are supposed to obey the indication given by the signal.
\item The level crossing control and its interaction with the routes is not modelled.
\item The different types of directional locking are not modelled. The directional locking is the mutual exclusion mechanism put in place to prevent head-to-head collisions.
\item The trains can postpone their start when in front of a signal at proceeding aspect but never stop afterwards. The train speed is not modelled.

\end{enumerate}

Our interlocking (SSI) is route based which means:
\begin{itemize}
\item A route must be successfully controlled by the controller before a train can run through the station.
\item The routes interact with the track side components (e.g.: points, signals).
\item The routes using shared resources (e.g.: points) make use of locking variables in order to prevent collisions.
\item The path followed by the train is based on the status of the track side components controlled by the routes.
\end{itemize}

\begin{figure}[ht]
\centering
\includegraphics[width=0.90\linewidth]{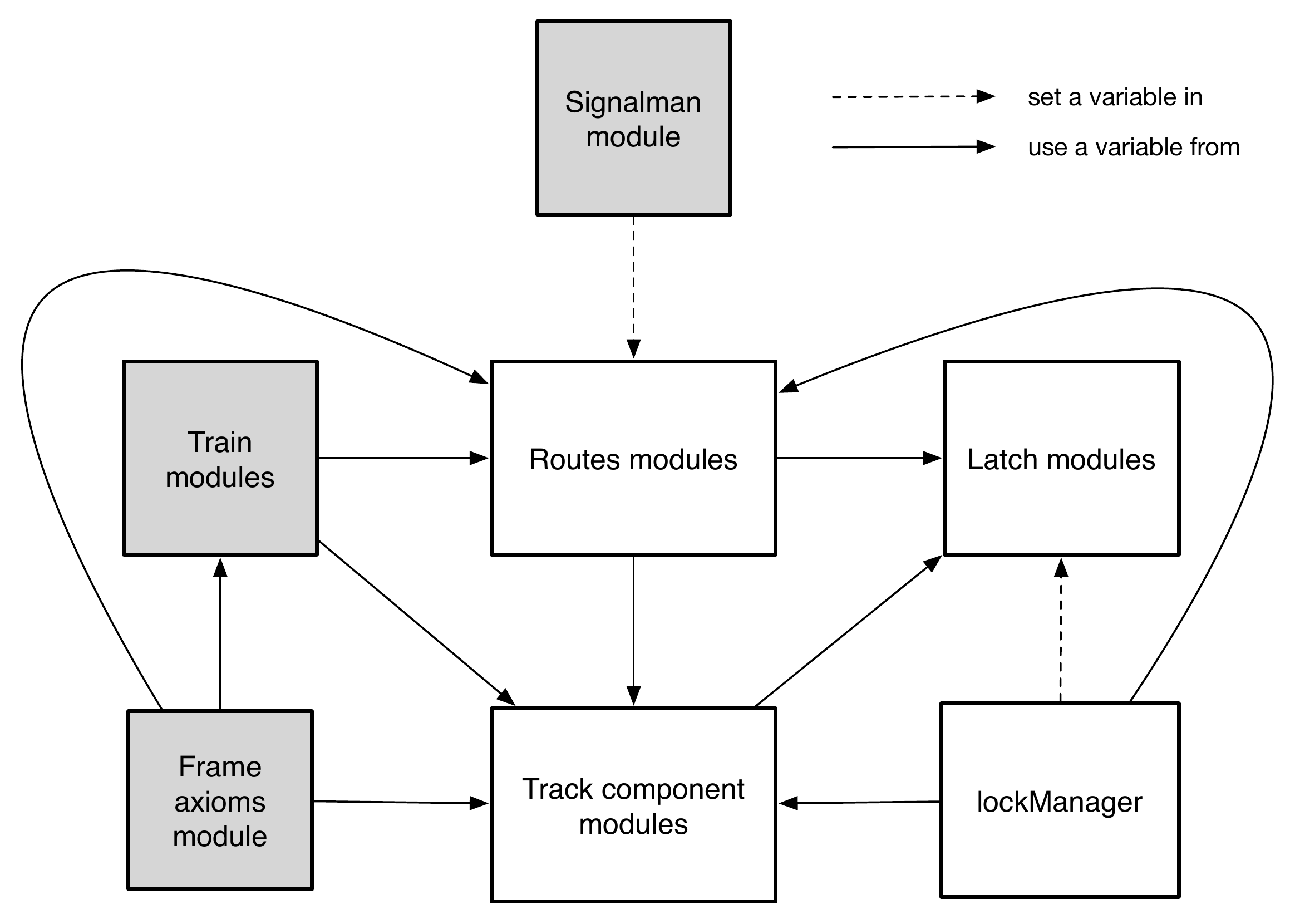} 
\caption{Modules view of the model.}
\label{fig:mView} 
\end{figure}

The Figure~\ref{fig:mView} shows how central the idea of route is in our model. The model is decomposed into NuSMV modules: the interlocking modules and the simulation modules. White modules model the interlocking software components while gray modules model components added to interact with the interlocking.

\begin{description}
\item[Latch modules:] The latches are the global variables shared by the route modules, the point modules, and the lock manager module. The UIR variable is an example of a latch. It is used to lock a point when it is part of a commanded route. The module acts as a record which state is updated by the lock manager module.
\\

\item[Track component modules:] The track modules represent a physical component controlled by the interlocking:

\begin{itemize}
	\item The track segments that hold the state of a track (occupied or clear).
	\item The points are commanded to the left, or the right position according to the route. 
\end{itemize}

\item[Lock manager module:] This module assume the task of locking and unlocking the subroutes and the points when a route is commanded and ran through by a train. This module is a straight emanation from the application data.
\\

\item[Controller module:] This module simulates the behaviour of a human commanding the routes. It also ensures that only one route command is issued per transition of the whole system. This module ensures that this behaviour is not violated.
\\
\item[Train module:] This module is used to simulate the movement of a train over the adjacent track segments forming the track layout of the station. The track layout is first encoded by mean of a DSL listing all the components like the signals, the track circuits, and the switches. Each component is linked to its siblings taking into account the train direction, and the position of the switches ahead of the train. The graph of the station is then built automatically and all the possible successive train positions are translated into the transitions of the NuSMV module allowing the train simulation.
\\

\item[Route modules:] The route lifecycle is described in Section \ref{compDesc}. The route modules are a straight translation of the application data from the SSI language to NuSMV. The state machine of a route includes the following states: idle, commanded, proved, and occupied by a train.
\\

\item[Frame axioms module:] This module performs three different tasks:

\begin{itemize}
	\item Changing the status of the track components according to the train movements.
	\item Triggering a wheel detector when a track segment is occupied.
	\item Updating the point position after a command.
\end{itemize}

This module depends on both the application data (routes) and the track layout (trains) to know when the actions must be done and what are the modifications to do.

\end{description}

Given that we want to verify the consistency between the application data and the real track layout, we have to consider a separate source for the application data and the layout. Therefore, unlike the other modules, the train module is not generated from the application data.

Put together, these modules constitute a model simulating the behaviour of an interlocking system as described in the application data and the behaviour of trains according to the track layout.
On this model, we can assert and automatically check safety properties with respect to the application data. These properties can be expressed on the state of the trains. For example, a collision occurs if two trains are both located on the same segment.  For instance, in Figure \ref{fig:layout}, such a collision will occur if the application data could allow routes R\_KM\_045 and R\_CM\_044 to be set together.

\section{Automatic translation of application data} 
\label{translation}

Among all the application data, only a subset is necessary to verify the security of an interlocking system.
The rest is either not related to the security or abstracted in our model.
Let us now describe the application data used in our model.


Each point can move under a set of conditions. Listing~\ref{pointConditions} shows how these conditions are represented in the SSI code
for a particular point. There are two positions for a point: normal  and reverse.\footnote{Normal stands for left and reverse for right.}
Here, the point P\_01AM can be set in a normal position (P\_01AM\textbf{N}) only if it is is free to move (U\_IR(01AM) f).
There is a similar rule for the reverse position (P\_01AM\textbf{R}) .

\begin{lstlisting}[style=SSI, caption=SSI code: Conditions allowing a point to move., label=pointConditions]
*P_01AMN  U_IR(01AM) f /* condition for normal position */ 
*P_01AMR  U_IR(01AM) f /* condition for reverse position */ 

\end{lstlisting}


Each route has a set of conditions under which the route request can be granted, and a set of actions that have to be done to fulfil the request.  For example, Listing~\ref{routeRequest} states that the route from Signal CM to Track 044 can only be set if it is not already set (line 2) and if the points are free to be commanded and moved to a certain position (lines 3 and 4). The resulting actions are the setting of the route (line 6), the command of the points (lines 7 and 8) and the locking of the points (line 9).
The route and the components requested can be seen on Figure~\ref{fig:layout}.
\\

\begin{lstlisting}[style=SSI, caption=SSI code: Request for setting a route.,  label=routeRequest]
*Q_R(CM_044) /* Request for the route CM_044 */
	if		R_CM_044 xs, 
			P_01AM cfr, P_02BM cfr, P_04AM cfr, 
			P_04BM cfr, P_01BM cfr, P_02AM cfr, 
			U_IR(01AM) f, U_IR(02BM) f, U_IR(04BM) f 
	then  R_CM_044 s
			P_01AM cr, P_02BM cr, P_04AM cr, 
			P_04BM cr, P_01BM cr, P_02AM cr, 
			U_IR(01AM) l, U_IR(02BM) l, U_IR(04BM) l 
\end{lstlisting}


After being locked for a route, the different track components must be freed.
According to Listing~\ref{subrouteFreeing}, the subroute U\_04M\_CM can only be freed if the subroute U\_07M\_04M is free  and if the track T\_04BM is clear. There is a set of similar rules for the liberation of the other subroutes and for other components.
\\

\begin{lstlisting}[style=SSI, caption=SSI code: Freeing a subroute.,  label=subrouteFreeing]
U_04M_CM f if U_07M_04M f, T_04BM c 
\end{lstlisting}

All these data are used to build the NuSMV model.
Each interlocking system has its own application data. 
In other words, we need to build a particular model for each interlocking system.
To overcome this issue, we designed a translator which automatically parses the application data and generates the NuSMV model.
In this way,  we can directly obtain an executable model for each interlocking system. For instance, the NuSMV module corresponding to the route request of Listing~\ref{routeRequest} is showed on Listing~\ref{lst:RouteSMV}.

\begin{lstlisting}[style=SMV, caption=A route command in the NuSMV model, label=lst:RouteSMV]
MODULE R_CM_044(mainP)
VAR
cmd : boolean; 
state : {s, xs}; -- set or unset
ASSIGN
init(cmd) := FALSE;
init(state) := xs;
next(state) := 
	case
		-- conditions to set the route
		state = xs & cmd &  -- route not already set
		mainP.UIR_04BM.st = f & -- track component is free
		mainP.UIR_02BM.st = f & 
		mainP.UIR_01AM.st = f &
		mainP.P_01AM.cfr & -- free to go to reverse position
		mainP.P_02BM.cfr & 
		mainP.P_04AM.cfr & 
		mainP.P_04BM.cfr & 
		mainP.P_01BM.cfr &
		mainP.P_02AM.cfr : s; 
		-- conditions to release the route
		(...)
	esac;
(...)
\end{lstlisting}

As we can see, this module contains the necessary conditions  to set the route. 
Let us note that the examples presented here do not show all the application data used in our model; other structures such as the train detectors are also used.

\section{Safety properties}
\label{safProp}


The safety properties verified on our model are expressed by mean of invariants (properties that are true in any state of the system) and CTL (Computation Tree Logic) formulas. A first set of properties is used to verify that the interaction between the interlocking and the train never ends up in an unsafe sequence causing train collisions or derailments. A second set of properties is used to detect which are the errors in the application data leading to an unsafe behaviour of the interlocking.  

Listing~\ref{lst:prop1} shows a sample of properties covering these sets, for route R\_CM\_044 of the Nam\^eche model.

\begin{lstlisting}[style=SMV, caption=Safety properties for the Nam\^eche model, label=lst:prop1]
INVARSPEC ! (train_1.front = derailed | train_2.front = derailed)
INVARSPEC ! (train_1.front = train_2.front)
INVARSPEC ! ((train_1.T_01AM | train_2.T_01AM) & P_01AM.willMove)
INVARSPEC ! (R_CM_044.st =  s & R_KM_045.st = s)
INVARSPEC ! (U_CM_04M.st = l & U_04M_CM.st = l)
CTLSPEC AG (T_04BM.st = o & TRP_CM.krc = s -> AX (!R_CM_043.L_CS & !R_CM_044.L_CS))
INVARSPEC ! (R_CM_044.st = s & U_CM_04M.st = l  & U_04M_07M.st = f)
INVARSPEC ! (UIR_01AM.st = l & P_01AM.willMove)
INVARSPEC (P_01AM.cmd = P_01BM.cmd)
INVARSPEC (R_CM_044.L_CS -> (T_04BM.st = c & T_02BM.st = c & T_01AM.st = c))	
\end{lstlisting}

\textbf{Safe interaction between the interlocking and the train} This first set embeds the properties verifying that an active simulation of two trains running through the network controlled by the interlocking does not lead to unsafe situations.

\begin{itemize}

\item Line 1: Trains never derail.  Trains derail when entering a point not set in a position allowing the train to continue its path.
\item Line 2: Trains never collide. The property is expressed by stating that the heads of the trains cannot occupy the same position at the same time.
\item Line 3: A point (P\_01AM) is not allowed to move when its home track-circuit is occupied.

\end{itemize}

\textbf{Application data correctness} A mistake or an omission in the application data causes the violation of some properties in the first set (e.g. a train collision). However, given a trace leading to a train collision, the identification of which part of the application data is faulty is not trivial. Each property of this second set concerns a route, a part of a route, or a point. As a result, finding the faulty part of the application data in case of violation is easier, as explained hereunder.

\begin{itemize}

\item Line 4: Incompatible routes (R\_KM\_045 and R\_CM\_044) are never enabled at the same time.
\item Line 5: Subroutes in opposite directions (U\_CM\_04M and U\_04M\_CM) are never locked at the same time.
\item Line 6: The origin signal of a route (R\_CM\_043 or R\_CM\_044) immediately goes back to danger after the train has started to run through it.  The formula states that in all states (AG) where the train is occupying the track-circuit and has activated the passage detector, the signal is closed in the next state for all possible executions (AX).  
\item Line 7: The subroutes are released in correct order.  In this case, subroute U\_04M\_07M is not released before subroute U\_CM\_04M.
\item Line 8: A point (P\_01AM) will not move when its locking variable (UIR\_01AM) is set.
\item Line 9: Connecting points (P0\_1AM and P\_01BM) are always commanded to the same position.
\item Line 10: In order to clear the origin signal of a route (R\_CM\_044.L\_CS), all the track circuits must be clear (not occupied by a train).

\end{itemize}

\section{Verification of properties}
\label{verification}

The model of Nam\^eche station comprises 14 routes, 7 points, and 7 track-circuits for which 132 invariants and 7 CTL formulas were written. These formulas were written manually based on the track layout of Figure~\ref{fig:layout} for the sake of independence from the application data. 

In order to test our model and the adequacy of the properties used on it, we have introduced errors in the application data. The violations were successfully detected and traces were generated. For instance, assigning a wrong locking variable to Point P\_01AM can lead to a situation where it is moved under a running train, failing Property~3 in Listing~\ref{lst:prop1}.
As another example, a missing locking of subroute U\_CM\_04M leads to a collision between two trains, one going from Signal CM to Point P\_04BM and one going from P\_04BM to CM, detected by Property 2 in Listing~\ref{lst:prop1}.

If we only consider the invariants, the verification with NuSMV takes less than fifteen minutes. However, with CTL formulas it takes few days to complete. To overcome this problem, we implemented custom model-checking algorithms with PyNuSMV.

First, our model includes fairness constraints\footnote{In the framework of CTL, fairness constraints are sets of states that must be met infinitely often along executions of interest~\cite{Clarke1999}.} used to force the trains to eventually progress on the tracks. The trains are modelled such that they can choose to wait at a green signal for an arbitrary long time; fairness constraints are specified to ensure that we consider only executions of the model along which the trains eventually choose to cross a green signal.   NuSMV performs model checking of CTL with fairness, which incurs additional computations of fair states and in the verification of some (liveness) formulas.
However, the CTL formulas verified on the model are not impacted by the presence of these fairness constraints because all reachable states of the model are fair and no liveness formula is verified. Thus, to accelerate the verification, standard BDD-based algorithms for CTL without fairness constraints have been used within PyNuSMV instead of those of NuSMV (see for example~\cite{Clarke1999} for more information on these standard algorithms). 

Furthermore, these CTL formulas follow the pattern $AG \phi$, where $\phi$ is a CTL sub-formula, expressing that $\phi$ is true in all reachable states. NuSMV verifies such formulas by starting from the BDD representing the set of states satisfying $\phi$ and performing a backward traversal of the system to accumulate the states satisfying the property; then NuSMV compares these states with the BDD of initial states. Nevertheless, this particular case can be improved by checking that all the reachable states satisfy $\phi$. This comparison is performed by comparing the BDD of the states satisfying $\phi$ and the BDD of reachable states, requiring one operation on the two BDDs, instead of the fix-point computation performed by the standard algorithm implemented in NuSMV. Note that this custom approach needs to compute the BDD representing the set of reachable states, but this BDD must be computed to verify the invariants. This BDD is computed by NuSMV itself by performing a standard forward traversal of the model.

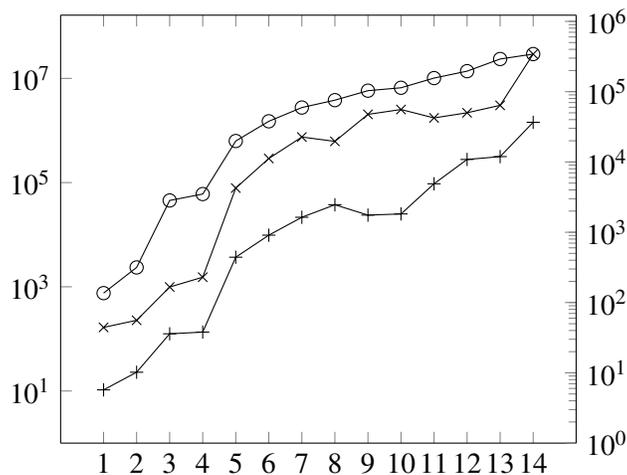
\begin{figure}[!ht]
	\centering
	\begin{tikzpicture}
	\pgfplotstableread{data-time-nusmv.txt}\nusmvdata;
	\pgfplotstableread{data-time-pynusmv.txt}\pynusmvdata;
	\pgfplotstableread{data-states.txt}\statesdata;
	
	\begin{semilogyaxis}[axis y line*=left, ymin=1,
								xtick=data]
	
	\addplot[mark=o, mark size=2.5] table {\statesdata};
	\label{plot:states}
	
	\end{semilogyaxis}

	\begin{semilogyaxis}[axis y line*=right,
							  	axis x line=none,
								xtick=data,
							  	ymin=1]

	\addplot[mark=x, mark size=2.5] table{\nusmvdata};
	\label{plot:nusmv}
	
	\addplot[mark=+, mark size=2.5] table{\pynusmvdata};
	\label{plot:pynusmv}
	
	\end{semilogyaxis}
			
	\end{tikzpicture}
	\vspace{-5mm}
	\caption{Evolution of number of reachable states ($\circ$) on the left y-axis and
	verification time for NuSMV ($\times$) and PyNuSMV ($+$) on the right y-axis (in seconds) in terms of number of considered routes.
	}
	\label{figure:verif-time}
\end{figure}

Thanks to these custom algorithms, it has been possible to verify the 7 CTL formulas of the previous section within 10 hours, while NuSMV took about 100 hours to verify them. Figure~\ref{figure:verif-time} recaps the execution time in function of the number of routes considered.



In order to assess the performance of the custom model-checking algorithms, we have also compared the verification time needed by both tools for models with a reduced number of routes. As we can see in Figure~\ref{figure:verif-time}, PyNuSMV reduces drastically the execution time when many routes are involved. More precisely, Figure~\ref{figure:verif-time} shows the time needed to verify the properties of the previous section with NuSMV 
and PyNuSMV 
when the number of considered routes increases. It shows that NuSMV and PyNuSMV behave in similar ways, but PyNuSMV gains an order of magnitude by using custom algorithms.

PyNuSMV can also be used to extract from the model the set of compatible routes. We say that a route is compatible with another route if they can be commanded at the same time, that is, if a train can pass through the first one while another train passes through the second one. Table~\ref{nameche-compatibility-table} shows the compatibility table for Nam\^eche (Figure~\ref{fig:layout}). It shows, for example, that Route R\_CM\_043 and Route R\_CXM\_044  are compatible while Routes R\_KM\_045 and R\_KM\_046 are not. This means that the interlocking system works such that whenever Route R\_KM\_045 is set, Route R\_KM\_046  cannot be commanded.

\begin{table}[h!t]
\centering
\begin{tabular}{l|cccccccccccccc}
    & \rotatebox{90}{R\_CM\_044} & \rotatebox{90}{R\_CM\_043} & \rotatebox{90}{R\_CXM\_044} & \rotatebox{90}{R\_CXM\_043} & \rotatebox{90}{R\_FM\_424} & \rotatebox{90}{R\_GM\_424} & \rotatebox{90}{R\_GM\_044} & \rotatebox{90}{R\_GYM\_423} & \rotatebox{90}{R\_GYM\_425} & \rotatebox{90}{R\_KM\_045} & \rotatebox{90}{R\_KM\_046} & \rotatebox{90}{R\_KM\_423} & \rotatebox{90}{R\_KXM\_045} & \rotatebox{90}{R\_KXM\_046} \\
\hline
R\_CM\_044  & \cellcolor{gray!60} &    &    &    &  V &  V &    &  V &  V &     &     &     &     &     \\
R\_CM\_043 & \cellcolor{gray!60} & \cellcolor{gray!60} &  V &    &  V &  V &  V &  V &  V &     &   V &   V &     &     \\
R\_CXM\_044  & \cellcolor{gray!60} & \cellcolor{gray!60} & \cellcolor{gray!60} &    &  V &  V &    &  V &  V &     &     &     &   V &     \\
R\_CXM\_043  & \cellcolor{gray!60} & \cellcolor{gray!60} & \cellcolor{gray!60} & \cellcolor{gray!60} &  V &  V &  V &  V &  V &     &     &   V &     &     \\
R\_FM\_424  & \cellcolor{gray!60} & \cellcolor{gray!60} & \cellcolor{gray!60} & \cellcolor{gray!60} & \cellcolor{gray!60} &    &  V &    &    &   V &   V &   V &   V &   V \\
R\_GM\_424  & \cellcolor{gray!60} & \cellcolor{gray!60} & \cellcolor{gray!60} & \cellcolor{gray!60} & \cellcolor{gray!60} & \cellcolor{gray!60} &    &    &    &   V &   V &     &   V &   V \\
R\_GM\_044  & \cellcolor{gray!60} & \cellcolor{gray!60} & \cellcolor{gray!60} & \cellcolor{gray!60} & \cellcolor{gray!60} & \cellcolor{gray!60} & \cellcolor{gray!60} &    &  V &     &     &     &   V &   V \\
R\_GYM\_423  & \cellcolor{gray!60} & \cellcolor{gray!60} & \cellcolor{gray!60} & \cellcolor{gray!60} & \cellcolor{gray!60} & \cellcolor{gray!60} & \cellcolor{gray!60} & \cellcolor{gray!60} &    &   V &   V &     &   V &   V \\
R\_GYM\_425  & \cellcolor{gray!60} & \cellcolor{gray!60} & \cellcolor{gray!60} & \cellcolor{gray!60} & \cellcolor{gray!60} & \cellcolor{gray!60} & \cellcolor{gray!60} & \cellcolor{gray!60} & \cellcolor{gray!60} &   V &   V &   V &   V &   V \\
R\_KM\_045 & \cellcolor{gray!60} & \cellcolor{gray!60} & \cellcolor{gray!60} & \cellcolor{gray!60} & \cellcolor{gray!60} & \cellcolor{gray!60} & \cellcolor{gray!60} & \cellcolor{gray!60} & \cellcolor{gray!60} & \cellcolor{gray!60} &     &     &     &     \\
R\_KM\_046 & \cellcolor{gray!60} & \cellcolor{gray!60} & \cellcolor{gray!60} & \cellcolor{gray!60} & \cellcolor{gray!60} & \cellcolor{gray!60} & \cellcolor{gray!60} & \cellcolor{gray!60} & \cellcolor{gray!60} & \cellcolor{gray!60} & \cellcolor{gray!60} &     &   V &     \\
R\_KM\_423 & \cellcolor{gray!60} & \cellcolor{gray!60} & \cellcolor{gray!60} & \cellcolor{gray!60} & \cellcolor{gray!60} & \cellcolor{gray!60} & \cellcolor{gray!60} & \cellcolor{gray!60} & \cellcolor{gray!60} & \cellcolor{gray!60} & \cellcolor{gray!60} & \cellcolor{gray!60} &   V &   V \\
R\_KXM\_045& \cellcolor{gray!60} & \cellcolor{gray!60} & \cellcolor{gray!60} & \cellcolor{gray!60} & \cellcolor{gray!60} & \cellcolor{gray!60} & \cellcolor{gray!60} & \cellcolor{gray!60} & \cellcolor{gray!60} & \cellcolor{gray!60} & \cellcolor{gray!60} & \cellcolor{gray!60} & \cellcolor{gray!60} &     \\
R\_KXM\_046& \cellcolor{gray!60} & \cellcolor{gray!60} & \cellcolor{gray!60} & \cellcolor{gray!60} & \cellcolor{gray!60} & \cellcolor{gray!60} & \cellcolor{gray!60} & \cellcolor{gray!60} & \cellcolor{gray!60} & \cellcolor{gray!60} & \cellcolor{gray!60} & \cellcolor{gray!60} & \cellcolor{gray!60} & \cellcolor{gray!60}
\end{tabular}

\caption{The compatibility table of the station of  \emph{Nam\^eche}.}
\label{nameche-compatibility-table}
\end{table}

The value $V$ in the table means that the corresponding routes are compatible, otherwise they are not compatible. Given that the table is symmetric, only the top half is presented.
Thanks to PyNuSMV, such a compatibility table can be extracted by inspection of the set of reachable states.
This table can then be used to check that the routes that should not be compatible are not, giving essential information on the application data under interest.
Compatibility sets of more than two routes can be produced in the same way: for the Nam\^eche station, 32 sets of three compatible routes exist (e.g. Routes R\_CM\_043, R\_GYM\_425 and R\_KM\_423 can be commanded at the same time), but no set of more than three compatible routes exists.

\section{Related work}
\label{relWork}

In \cite{SSIgdlSMV}, Huber and King demonstrates how five vital safety properties can be verified automatically on SSI application data. They implemented a model checker for Geographic Data by replacing the parser and compiler of NuSMV. The resulting tool, gdlSMV, directly reads Geographic Data and builds a corresponding representation on which model checking is performed using NuSMV's symbolic model checking algorithms. In \cite{Iran}, Mirabadi and Yazdi also use the NuSMV model checker and implement a control table verifier that analyses the contents of control table besides the safe train movement conditions and checks for any conflicting settings in the table. In \cite{WIN1}, Winter and Robinson modelled the interlocking by means of the formal notation ASM that are more readable. The formal model is translated in NuSMV code and the Safety requirements are expressed in CTL.
\par In \cite{CSPB2,CSPB3,CSPB5}, Moller, Nga Nguyen, Roggenbach, Schneider and Treharne propose to combine the state-based and the event-based (a train passing) approaches by using CSP$\|$B. The overall specification combines two communicating models, one made of CSP process descriptions and one made of a collection of B machines. They also propose the OnTrack tool-set that automates workflows for railway verification, starting with graphical scheme plans and finishing with automatically generated formal models set up for verification. In \cite{Work2013_8}, Abo and Voisin explain how Systerel uses the B language and the OVADO tool to verify large interlocking application data set.
\par In \cite{genView}, Fantechi, Fokkink and Morzenti give an overview of the trend in railway interlocking verification. 

\par Compared with the previous works, we presented here an unified approach aiming to verify completely the safety of an interlocking system using model checking. More concretely, our approach has the following features:

\begin{itemize}
\item A verification of the correctness of the application data.
\item A verification of their consistency with the track layout.
\item An automatic generation of the models used for the the verification from the application data.
\item A Domain Specific Language used to easily encode a track layout into the models.
\end{itemize}

Taken separately, such features have already been discussed and considered. But to the best of our knowledge, there is no work that merges all of them into a single framework.

\section{Conclusions and future work}
\label{futWork}

In this paper, we have explained how we built a model of a railway interlocking in order to verify the correctness of its application data. We have explained how each module of the model can be automatically generated based on the application data by our generator. We have given the list of safety properties that were verified on our model. Those safety properties are designed to cover the tests and verifications that are currently performed manually on the application data. We have shown that the verification of those properties by a model checker brings improvement in the safety and in the efficiency of the verification process ruling the validation of the application data. Finally, we have shown that the verification of large amount of properties (137) is feasible on a realistic size interlocking by mean of custom algorithms based on PyNuSMV.

In our future work, we will focus on the automatic generation of the safety properties based on the track layout. The aim is to use a railway description language such as RailML~\cite{railML} from which we can generate the safety rules. Furthermore, the properties we verify on the model have been limited to invariants and safety properties; verifying liveness properties such as \textit{any train entering a route will eventually leave it} needs more effort, and further work is needed to efficiently verify such properties.
Besides, until now the model is only designed to verify single interlockings. This assumption holds for relatively small stations such as our case study but is not true for larger stations where several interlockings communicate together. 
The next step will be to extend the model in order to embed the verification of a set of communicating interlockings.
Finally, as we plan to verify larger interlocking systems, we will have to work on increasing the efficiency of model-checking algorithms.

\bibliography{./reducedBIB}

\end{document}